\documentclass[10pt,aps,pra,twocolumn,superscriptaddress,floatfix]{revtex4-1}
\pdfoutput=1
\usepackage[utf8]{inputenc}
\usepackage[T1]{fontenc}
\usepackage{amsmath}
\usepackage{amssymb}
\usepackage{amsfonts}
\usepackage{bm}
\usepackage{bbm}
\usepackage{epsfig}
\usepackage{grffile}
\usepackage{graphics}

\usepackage{numprint}

\usepackage[usenames,dvipsnames]{color}
\definecolor{dblue}{rgb}{0,0.1,.6}

\usepackage[colorlinks=true,citecolor=dblue,linkcolor=dblue,urlcolor=dblue]{hyperref}
\usepackage[all]{hypcap}

\newcommand{\id}{\mathbbm{1}}
\newcommand{\bra}{\langle}
\newcommand{\ket}{\rangle}
\renewcommand{\vec}[1]{{\boldsymbol{#1}}}

\newcommand{\mc}[1]{\mathcal{#1}}

\newcommand{\hH}{\hat{H}}
\newcommand{\hh}{\hat{h}}
\newcommand{\hT}{\hat{T}}
\newcommand{\B}{\mc{B}}
\renewcommand{\O}{\mc{O}}
\newcommand{\CC}{\mathbb{C}}
\newcommand{\eff}{\text{eff}}

\newcommand{\duke}  {Department of Physics, Duke University, Durham, North Carolina 27708, USA}
\newcommand{\ustc}{School of the Gifted Young, University of Science and Technology of China, Hefei, Anhui 230026, China}

\begin{document}

\title{Tensor Network States with Low-Rank Tensors}
\author{Hao Chen}
\affiliation{\duke}
\affiliation{\ustc}
\author{Thomas Barthel}
\affiliation{\duke}

\begin{abstract}
Tensor networks are used to efficiently approximate states of strongly-correlated quantum many-body systems. More generally, tensor network approximations may allow to reduce the costs for operating on an order-$N$ tensor from exponential to polynomial in $N$, and this has become a popular approach for machine learning. We introduce the idea of imposing low-rank constraints on the tensors that compose the tensor network. With this modification, the time and space complexities for the network optimization can be substantially reduced while maintaining high accuracy.

We detail this idea for tree tensor network states (TTNS) and projected entangled-pair states. Simulations of spin models on Cayley trees with low-rank TTNS exemplify the effect of rank constraints on the expressive power. We find that choosing the tensor rank $r$ to be on the order of the bond dimension $m$, is sufficient to obtain high-accuracy groundstate approximations and to substantially outperform standard TTNS computations. Thus low-rank tensor networks are a promising route for the simulation of quantum matter and machine learning on large data sets.
\end{abstract}

\date{May 21, 2022}

\maketitle

\section{Introduction}
Tensor network states (TNS) have gained tremendous success in quantum many-body physics. Consider a lattice system with $N$ sites (or orbitals), each associated with a site Hilbert space of dimension $d$. In many problems, TNS resolve the curse of dimensionality associated with the exponential growth of the Hilbert space dimension $d^N$ with the system size. They approximate quantum states $|\Psi\ket$ by a network of partially contracted tensors. The tensors may carry physical indices $\sigma_i=1,\dotsc,d$ that label site basis states and additional bond indices of dimension $m$, which are contracted with corresponding indices of other tensors \cite{Orus2014-349}.
If the network structure is well-aligned with the entanglement structure of the system, good TNS approximations of $|\Psi\ket$ can be achieved with only $\O(N)$ parameters.
The simplest type of TNS are matrix product states (MPS) \cite{Fannes1992-144,White1992-11,Rommer1997,PerezGarcia2007-7,Schollwoeck2011-326}, which are most suitable for one-dimensional systems and lie at the heart of the density-matrix renormalization group (DMRG) algorithm \cite{White1992-11,Rommer1997}. Further types of TNS that are most useful for higher-dimensional systems are tree tensor network states (TTNS) \cite{Shi2006-74,Murg2010-82,Nakatani2013-138}, the multiscale entanglement renormalization ansatz (MERA) \cite{Vidal-2005-12,Vidal2006}, and projected entangled-pair states (PEPS) \cite{Niggemann1997-104,Nishino2000-575,Verstraete2004-7,Verstraete2006-96}. Beyond applications in physics, tensor networks recently also spurred great interest in the machine learning community, where tensor networks can be used for both supervised and unsupervised learning \cite{Cohen2016-29,Stoudenmire2016-29,Novikov2016_05,Han2018-8,Liu2019-21,Stoudenmire2018-3,Grant2018-4,Huggins2019-4,Cheng2019-99,Efthymiou2019_06,Liu2021_08}.

While we have gained many important insights through tensor network studies, the applicability to complex problems is limited by the computation costs. Tensor network contraction costs scale as $\O(m^{z+1})$ in the bond dimension $m$ for TTNS on a graph with coordination number $z$, as $\O(m^{7\dots 9})$ for one-dimensional (1D) MERA \cite{Evenbly2013}, as $\O(m^{10\dots 12})$ for 2D PEPS \cite{Murg2007-75,Jordan2008-101,Orus2009_05,Corboz2016-94}, and as $\O(m^{16\dots 28})$ for 2D MERA \cite{Cincio2008-100,Evenbly2009-102}. Hence, practicable bond dimensions $m$ are usually rather small, which limits the approximation accuracy.

In this work, we show how the number of tensor network parameters and the time complexity can be reduced by working with tensors of limited canonical polyadic (CP) rank, i.e., networks of \textit{low-rank tensors}. The idea is borrowed from the canonical polyadic decomposition \cite{Hitchcock1927-6,Carroll1970-35,Harshman1970-16,Kolda2009-51}, which has a wide range of applications in data analysis \cite{Kolda2009-51}. For concreteness, we will describe the approach in detail for TTNS (Secs.~\ref{sec:TTNS} and \ref{sec:lrTTNS}) and study how the tensor-rank constraints affect the variational power of the networks (Sec.~\ref{sec:benchmark}). Moreover, we describe a low-rank adaptation for 2D PEPS with open boundary conditions and explicitly specify the contraction order for evaluating norms, expectation values, and gradients, reducing the computational complexity of PEPS (Sec.~\ref{sec:lrPEPS}).

\section{Low-rank tensor networks}\label{sec:lrTNS}
\begin{figure*}[t]
	\includegraphics[width=\textwidth]{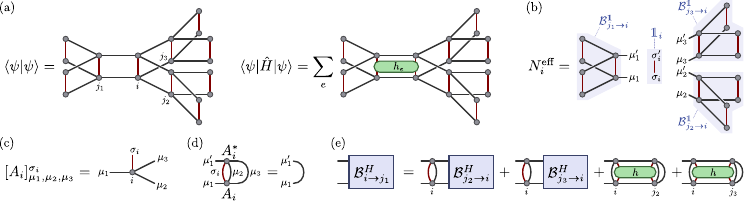}
	\caption{\textbf{TTNS and contractions for full-rank tensors.} (a) Diagrammatic representations for a squared norm $\bra\psi|\psi\ket$ and the energy expectation value $\bra\psi|\hH|\psi\ket$ of a TTNS with full-rank tensors as indicated in (c). (b) The norm matrix $N^\eff_i$ is obtained by removing tensors $A_i$ and $A^*_i$ from the network for $\bra\psi|\psi\ket$. It is composed of the $m\times m$ identity branch matrices $\B^\id_{j_n\to i}$. When the tensors $A_i$ obey suitable orthonormality constraints \eqref{eq:ON-constr} as illustrated in (d), the $\B^\id_{j_n\to i}$ become identities. (e) Iterative computation of Hamiltonian branch matrices. With $\B^H_{j_2\to i}$ and $\B^H_{j_3\to i}$ given, $\B^H_{i\to j_1}$ is obtained through a contraction with tensor $A_i$ and interaction terms $\hh_e$. The orthonormality constraint \eqref{eq:ON-constr} is used to simplify this computation.}
	\label{fig:fTTNS_contraction}
\end{figure*}
The canonical polyadic decomposition is a generalization of the singular value decomposition to higher-order tensors, which
expresses a tensor as the sum of direct products of vectors
\begin{equation}\label{eq:CPD}
	 A = \sum_{k=1}^r a^{k}_1\otimes a^{k}_{2}\otimes \dotsb \otimes a^{k}_{z}\ \in\ \CC^{m_1 \times \dotsb \times m_z},
\end{equation}
where $a^{k}_{e} \in \CC^{m_e}$. The minimal $r$ satisfying Eq.~\eqref{eq:CPD} is called the rank of $A$. The low-rank tensor approximation consists in limiting the maximum tensor rank $r$.

In a low-rank TNS, we choose every tensor to be of the form \eqref{eq:CPD}. Hence, the number of parameters per tensor drops from $\O(m_1\times\dots\times m_z)$ to $\O(r(m_1+\dots m_z))$, which may reduce the expressiveness of the TNS as a variational ansatz, but also substantially reduce computation costs. In particular, the time complexity for tensor contractions can be reduced by separately contracting the rank-one tensors $a^{k}_e$, and summing over $k=1,\dotsc,r$ in the end. We will see this in detail for TTNS and PEPS.

\section{Tree tensor network states}\label{sec:TTNS}
Let us first recall the traditional TTNS with full-rank tensors (fTTNS) \cite{Shi2006-74}. Consider a Cayley tree with $N$ physical sites and coordination number $z$. To each vertex $i$ with nearest neighbors $j_1,\dotsc,j_z$, we assign a tensor $[A_i]^\sigma_{\mu_1\dotsc,\mu_z}$ with one virtual index $\mu_n$ of dimension $m$ for each edge to a nearest neighbor and one physical index $\sigma$ of dimension $d$. The tensors for the leaves of the tree carry only one virtual and one physical index. The \mbox{fTTNS} is obtained by contracting the virtual indices for all edges (denoted by $\operatorname{Contr}$)
\begin{equation}
	|\psi\ket = \sum_{\sigma_1,\dotsc, \sigma_N} \operatorname{Contr}(A^{\sigma_1}_1\dotsb A^{\sigma_N}_N)|\sigma_1,\dotsc, \sigma_N\ket,
\end{equation}
where $\{|\sigma_i\ket\,|\,\sigma_i=1,\dotsc,d\}$ is an orthonormal basis for the Hilbert space of site $i$. For a given Hamiltonian $\hH$, the goal of the algorithm is to optimize the set of tensors $\{A_i\}$ such that the energy $E=\bra\psi|\hH|\psi\ket/\bra\psi|\psi\ket$ is minimal. This is equivalent to minimizing the functional
\begin{equation}
	F = \bra\psi|\hH|\psi\ket - E\bra\psi|\psi\ket.
\end{equation}

In the following, we assume that the Hamiltonian $\hH = \sum_e \hh_e$ only contains nearest-neighbor interactions, 
where $\hh_e$ is the two-site interaction term defined on edge $e$. Figure~\ref{fig:fTTNS_contraction}a gives graphical representations 
for $\bra\psi|\psi\ket$ and $\bra\psi|\hH|\psi\ket$.
Since the graph is acyclic, the optimization can be carried out very similarly to the DMRG algorithm \cite{White1992-11,Schollwoeck2011-326}. If we fix all the tensors except $A_i$, then $F$ is just a quadratic function of $A_i$, for which the minimization problem reduces to the generalized eigenvalue equation $H^\eff_i \vec{A}_i = E N^\eff_i \vec{A}_i$. Here we treat $A_i$ as a vector $\vec{A}_i\in\CC^{dm^z}$, and $H^\eff_i = \partial_{\vec{A}_i}\partial_{\vec{A}_i^*} \bra\psi|H|\psi\ket$ as well as $N^\eff_i = \partial_{\vec{A}_i}\partial_{\vec{A}_i^*}\bra\psi|\psi\ket$ as $dm^z\times dm^z$ matrices.

The diagrammatic representation of the effective norm matrix $N^\eff_i$ is given in Fig.~\ref{fig:fTTNS_contraction}b and shows that it is the tensor product $N^\eff_i=\id_i\otimes\B^\id_{j_1\to i}\otimes\dotsb\otimes \B^\id_{j_z\to i}$ of the identity on site $i$ and \emph{branch matrices} $\B^\id_{j_n\to i}$ that represent the identity acting on the $z$ branches obtained when removing vertex $i$ from the tree. These branch matrices are obtained by truncating the tensor network for $\bra\psi|\psi\ket$ at edges $j_n\to i$.
As in DMRG, we can impose orthonormality constraints on the tensors $A_{j\neq i}$ such that $\B^\id_{j_n\to i}=\id$ $\forall n$ and, hence, $N^\eff_i = \id$: First, designate vertex $i$ (the site to be optimized) as the root of the tree. We then have the notion of children and parents according to graph distances from the root. Second, starting from leaves, perform reduced RQ factorizations \cite{Golub1996} like
\begin{equation}
	[A_\ell]^{\sigma_\ell}_{\mu_1,\mu_2,\dotsc,\mu_z} =: \sum_{\mu} [R]_{\mu_1,\mu} [\tilde{A}_\ell]^{\sigma_\ell}_{\mu,\mu_2, \dotsc,\mu_z}
\end{equation}
for vertex $\ell$, where $\mu_1$ corresponds to the edge connecting vertex $\ell$ to its parent, and $\tilde{A}_\ell$ is an isometry satisfying
\begin{equation}\label{eq:ON-constr}
	\sum_{\sigma_\ell, \mu_2,\dotsc,\mu_z} [\tilde{A}_\ell]^{\sigma_\ell}_{\mu,\mu_2, \dotsc, \mu_z}[\tilde{A}^*_\ell]^{\sigma_\ell}_{\mu',\mu_2,\dotsc,\mu_z} = \delta_{\mu,\mu'}.
\end{equation}
See Fig.~\ref{fig:fTTNS_contraction}d. The orthonormalization of $A_\ell$ is achieved by replacing $A_\ell$ with $\tilde{A}_\ell$ and absorbing tensor $R$ into the parent of $\ell$. The cost of this operation is $\O(dm^{z+1})$. By performing this operation from the leaves up to the root $i$, we complete the orthonormalization with respect to $i$ such that $N^\eff_i$ is the identity by the virtue of Eq.~\eqref{eq:ON-constr}.

The computation of $H^\eff_i$ proceeds similarly. We first compute branch matrices $\B_{j_n\to i}^H$ that represent the Hamiltonian acting on the branches emanating from vertex $i$. This can be achieved by an iterative computation starting from the leaves. Figure~\ref{fig:fTTNS_contraction}e shows one step for a tree with $z=3$. The cost of the contraction is $\O(d^2m^{z+1})$. Once we have computed all branch matrices for site $i$, similar contractions yield the action of $H^\eff_i$ on $A_i$, i.e., the matrix-vector product needed in Krylov subspace methods for the solution of $H^\eff_i \vec{A}_i=E \vec{A}_i$. Thus, for an fTTNS with coordination number $z$, the cost for single-site DMRG scales as $\O(d^2m^{z+1})$. When moving from vertex to vertex in the DMRG-like alternating least-squares optimization, the orthonormalization center ($i$) and branch matrices can be updated locally without traversing the entire network.

Single-site DMRG is prone to getting stuck in local minima. Approaches to alleviate the problem are two-site DMRG \cite{White1996-77,Schollwoeck2011-326} and single-site DMRG with density-matrix perturbation \cite{White2005-72}. The corresponding algorithms on fTTNS turn out to have very high costs. In particular, we find the fTTNS time complexities to be
\begin{equation}\label{eq:fTTNScost}
	\O(d^2m^{z+1}),\ \
	\O(d^3m^{2z-1}),\ \ \text{and}\ \
	\O(d^3m^{3(z-1)})
\end{equation}
for single-site DMRG, two-site DMRG, and single-site DMRG with density-matrix perturbation, respectively.

\begin{figure}[t]
	\includegraphics[width=\columnwidth]{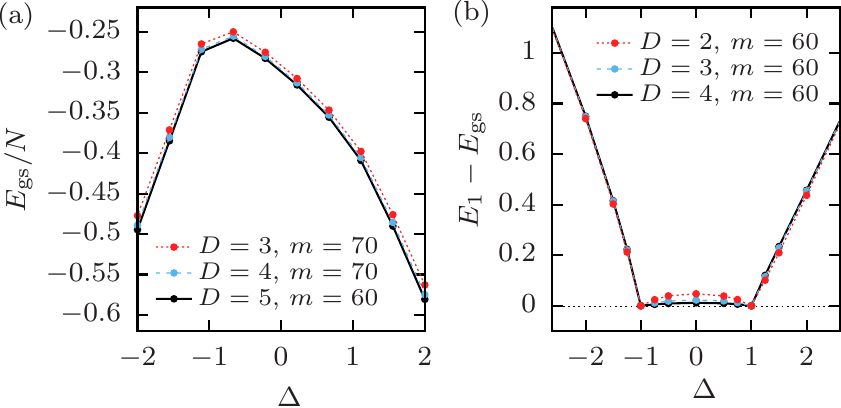}
	\caption{\textbf{fTTNS simulation for the XXZ model \eqref{eq:XXZ}.} (a) Groundstate energy per site $E_{\text{gs}}/N$ versus anisotropy $\Delta$ for $z=3$ and tree depths $D=3,4,5$. 
	(b) Energy gap for $z=3$.}
	\label{fig:fTTNS_results}
\end{figure}

As an example, consider the spin-$1/2$ XXZ model defined on a Cayley tree. The Hamiltonian of the model is
\begin{equation}\label{eq:XXZ}
	\hH = \sum_{\bra i,j \ket} \hat{S}^x_i\hat{S}^x_j + \hat{S}^y_i\hat{S}^y_j + \Delta \hat{S}^z_i\hat{S}^z_j,
\end{equation}
where the sum runs over all edges. We use fTTNS to obtain the approximate groundstate energies for coordination number $z=3$ with different tree depths as shown in Fig.~\ref{fig:fTTNS_results}a. Furthermore, excited states are computed by rerunning the energy minimization and orthogonalizing the new state to the previously determined ones (Fig.~\ref{fig:fTTNS_results}b). With increasing tree depth (system size), the energy gap approaches zero for $-1\leq\Delta\leq 1$ while it remains finite outside this region. The gapped ferromagnetic phase $\Delta < -1$ and antiferromagnetic N\'{e}el phase $\Delta > 1$ feature doubly degenerate ground states and spontaneous breaking of the $\mathbb{Z}_2$ symmetry. In the gapless phase $-1 < \Delta < 1$, the model has a unique ground state.

\section{Low-rank Tree Tensor Network States}\label{sec:lrTTNS}
\begin{figure}[t]
	\includegraphics[width=\columnwidth]{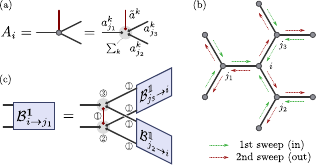}
	\caption{\textbf{Low-rank TTNS.} (a) lrTTNS are composed of rank-constrained tensors \eqref{eq:lrTTNS-Ai}. (b) Order for evaluating all branch matrices. The first sweep goes from leaves to an arbitrarily chosen root $i$. The second sweep goes from the root to the leaves. (c) The contraction order for computing the identity branch matrix $\B^\id_{i\to j_1}$. In Eq.~\eqref{eq:lrTTNS-Bid}, one can first form the $r\times r$ matrices $\bra \tilde{a}^{k}|\tilde{a}^{k'}\ket$ and $\bra a_{j_n}^{k}| \B^\id_{j_n\to i} |a_{j_n}^{k'}\ket$, then sum over $k$, and finally over $k'$.}
	\label{fig:lrTTNS}
\end{figure}

As the fTTNS time and space complexities scale exponentially with the lattice coordination number $z$ and with a correspondingly high power of the bond dimension $m$, the algorithm is infeasible for large $z$ and $m$. This problem can be solved by the low-rank TTNS (lrTTNS). Every tensor in the tree, except those for the leaves, then has the form
\begin{equation}\label{eq:lrTTNS-Ai}
	A_i = \sum_{k=1}^r \tilde{a}^{k} \otimes a_{j_1}^{k} \otimes a_{j_2}^{k} \otimes \dotsb \otimes a_{j_z}^{k}\ \in\ \CC^{d\times m^{\times z}},
\end{equation}
where $j_1,\dotsc,j_z$ are the nearest neighbors of site $i$, $\tilde{a}^{k} \in \CC^d$ and $a_{j_n}^{k} \in \CC^m$. A diagrammatic representation for such low-rank tensors is shown in Fig.~\ref{fig:lrTTNS}a. We will typically not impose orthogonality constraints on lrTTNS.

To obtain the norm, energy, and energy gradient of the state, we again need to evaluate the identity and Hamiltonian branch matrices $\B^\id_{j\to i}$ and $\B^H_{j\to i}$. The branch matrices for all edges (in both directions) can be obtained in two sweeps, first traversing from leaves to a root, and then from the root to leaves as indicated in Fig.~\ref{fig:lrTTNS}b. Figure~\ref{fig:lrTTNS}c specifies the contraction order for computing $\B^\id_{i\to j_1}$ from $\B^\id_{j_2\to i},\dotsc,\B^\id_{j_z\to i}$ and $A_i$ such that
\begin{equation}\label{eq:lrTTNS-Bid}
	\B^\id_{i\to j_1} = \sum_{k,k'} |a_{j_1}^{k'}\ket  \bra \tilde{a}^{k}|\tilde{a}^{k'}\ket
	  \prod_{n=2}^z\bra a_{j_n}^{k}| \B^\id_{j_n\to i} |a_{j_n}^{k'}\ket \bra a_{j_1}^{k}|,
\end{equation}
\begin{figure*}[t]
	\includegraphics[width=\textwidth]{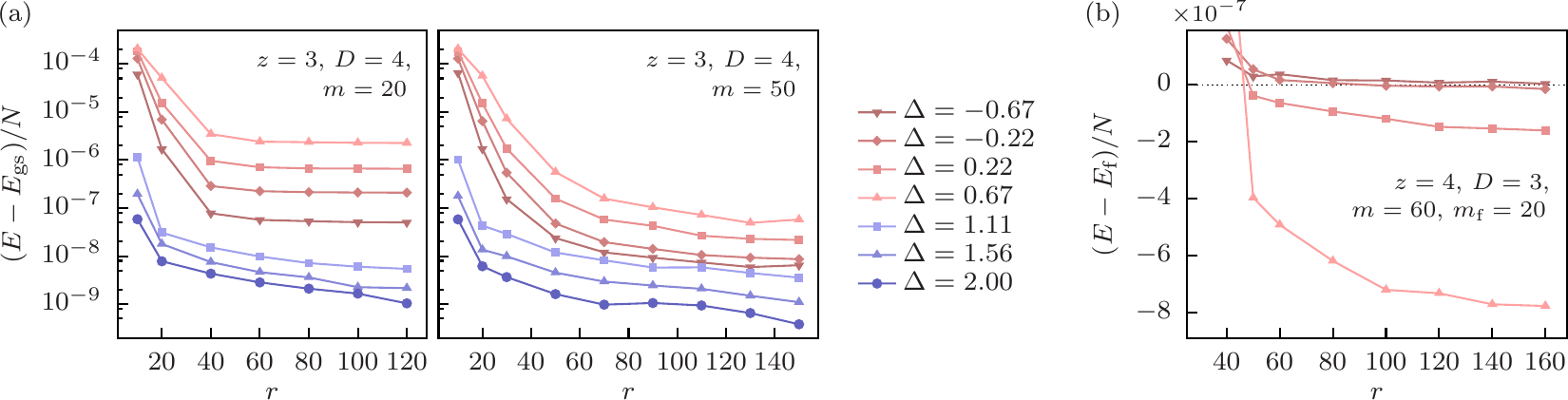}
	\caption{\textbf{Convergence and accuracy of low-rank TTNS for the XXZ model \eqref{eq:XXZ}.} 
	(a) The convergence of lrTTNS energies $E$ for a Cayley tree with coordination number $z=3$ and depth $D=4$ ($N=46$ sites) is shown as a function the tensor rank $r$. The shown energies are obtained after \numprint{80000} iterations.
	(b) For larger $z$, fTTNS simulations are very costly and we, hence, fixed their bond dimension to $m_\text{f}=20$. In contrast, lrTTNS have much lower computation costs, and one can afford larger $m$ and achieve lower energies. Even with $m=60$ and $r=160$, the number of parameters in the lrTTNS tensors is only $12\%$ of that in the $m=20$ fTTNS. The difference in the energy density $(E-E_{\text{f}})/N$ is shown for $z=4$ and depth $D=3$ ($N=53$) after \numprint{70000} iterations.}
	\label{fig:lrTTNS_convergence}
\end{figure*}
with $\bra a|\B|a'\ket:= \sum_{\mu,\mu'} a^*_{\mu} \B_{\mu,\mu'} a_{\mu'}$. The total cost for these contractions is $\O(z(r^2m+m^2r))$.
The Hamiltonian branch matrices $\B^H_{j\to i}$ can be evaluated with the same computational complexity. Given all branch matrices for an arbitrary site $i$, we can contract them with $A_i$ and $A_i^*$ (and Hamiltonian terms $\hh_e$ on the edges to the neighbors of site $i$) to get the squared norm $\bra\psi|\psi\ket$ and the expectation value $\bra\psi|H|\psi\ket$, and hence the energy of the state.

To find the ground state, one can use gradient-based algorithms like L-BFGS \cite{Nocedal2006,Liu1989-45} to minimize the energy, which is now feasible, given that the number of parameters per tensor is decreased from $m^zd $ to $r(zm+d)$. 
With the branch matrices for all edges computed, the gradient of the energy with respect to all rank-one components can be evaluated efficiently. Consider site $i$ with the tensor \eqref{eq:lrTTNS-Ai} as shown in Fig.~\ref{fig:lrTTNS}a. The derivatives of $\bra\psi|\psi\ket$ with respect to $\tilde{a}^{k*}$ and $a_{j_\nu}^{k*}$ are
\begin{equation*}
	\begin{aligned}
	\frac{\partial\bra\psi|\psi\ket}{\partial \tilde{a}^{k*}}
	  &= \sum_{k'=1}^{r} |\tilde{a}^{k'}\ket
	     \prod_{n=1}^z \bra a_{j_n}^{k}|\B^\id_{j_n\to i}|a_{j_n}^{k'}\ket\quad\text{and}\\
	\frac{\partial\bra\psi|\psi\ket}{\partial a_{j_\nu}^{k*}} 
	  &= \sum_{k'=1}^{r} \B^\id_{j_\nu\to i}|a_{j_\nu}^{k'}\ket \bra\tilde{a}^{k}|\tilde{a}^{k'}\ket 
	     \prod_{n\neq \nu} \bra a_{j_n}^{k}|\B^\id_{j_n\to i}|a_{j_n}^{k'}\ket.
	\end{aligned}
\end{equation*}
The derivatives of $\bra\psi|H|\psi\ket$ can be computed in the same manner. The computational cost of the above type of operations is again $\O(z(r^2m + m^2r))$. Therefore, the time complexity of all operations used in an lrTTNS optimization step is
\begin{equation}\label{eq:lrTTNScost}
	\O(z(r^2m+m^2r)),
\end{equation}
which makes it attractive for simulations with large bond dimensions $m$ as well as trees with large coordination numbers $z$.

\section{Benchmark simulations}\label{sec:benchmark}
To demonstrate and benchmark lrTTNS, we apply it to the spin-1/2 XXZ model \eqref{eq:XXZ}, minimizing the energy through the L-BFGS algorithm \cite{Nocedal2006,Liu1989-45}. There are various ways to avoid local minima. We use \emph{scanning}, where the anisotropy $\Delta$ is changed in small increments, traversing a few times the interval $\Delta\in[-2,2]$. Each step is initialized with the converged lrTTNS of the previous $\Delta$ value. A relatively small number of iterations (in our case 1000) at each point is sufficient to roughly locate the global minimum for all $\Delta$. Starting from scanning results, one can then perform more extensive optimizations at selected anisotropies to obtain precise groundstate approximations.

In the following, we only consider anisotropies $\Delta > -1$, because the ground state is a trivial ferromagnetic product state when $\Delta < -1$. In Fig~\ref{fig:lrTTNS_convergence}a, we use the deviation $\delta e := (E-E_{\text{gs}})/N$ of the groundstate energy density to quantify the expressiveness of the low-rank networks. Here, $E$ is the optimized lrTTNS energy for a given tensor rank $r$ and bond dimension $m=20,50$, and the exact groundstate energy $E_{\text{gs}}$ is approximated using fTTNS with a sufficiently large bond dimension $m_\text{f}>m$; here, $m_\text{f}=70$. The figure shows $\delta e$ as a function of $r$ for coordination number $z=3$, various $\Delta>-1$, and tree depths $D=4,5$. There is a fast reduction of $\delta e$ for $r<m$, and the speed of reduction decreases for $r\gtrsim m$. In the gapped N\'{e}el phase ($\Delta > 1$), lrTTNS can reach an accuracy $\delta e \lesssim\sim 10^{-8}$ for $r \sim m$. Ground states in the critical phase $|\Delta| < 1$ have substantially higher entanglement and one hence needs larger bond dimensions $m$ and tensor ranks $r$ to achieve high accuracy. In this case, $m=50$ and, again, $r\sim m$ are sufficient to reach $\delta e < 10^{-7}$.

The advantage of lrTTNS is more obvious in trees with larger coordination numbers. In Fig.~\ref{fig:lrTTNS_convergence}b, we compare the variational power of lrTTNS and fTTNS for $z=4$. The fTTNS bond dimension is now restricted to $m_\text{f}=20$ due to the high space complexity and time complexity \eqref{eq:fTTNScost} of fTTNS. However, the lrTTNS approach does not suffer this issue as its complexity \eqref{eq:lrTTNScost} is linear in $z$, which allows us to explore relatively large bond dimensions like $m=60$ for $z=4$. The figure only shows data for the critical phase $|\Delta| < 1$ as the energies for the gapped phases are already very accurate at considerably smaller $m$. Remarkably, we find that, although the number of parameters in the lrTTNS is significantly smaller than that in the employed fTTNS, lrTTNS can find substantially lower groundstate energies. For example, an lrTTNS with $m=60$ and $r=160$ has \numprint{38720} parameters per tensor, which is only $12\%$ of the number of parameters in an fTTNS with $m_\text{f}=20$, but it finds a more precise ground state as shown in Fig.~\ref{fig:lrTTNS_convergence}b. Not surprisingly, the improvement is largest for the most entangled state ($\Delta=0.67$).

\section{Low-rank Projected Entangled Pair States}\label{sec:lrPEPS}
The low-rank tensor network approach also works for PEPS \cite{Niggemann1997-104,Nishino2000-575,Verstraete2004-7,Verstraete2006-96,Jordan2008-101,Orus2009_05,Corboz2016-94}. PEPS are particularly suitable for the simulation of strongly-correlated 2D systems -- a class of systems that features some of the most exciting quantum phenomena like spin liquids \cite{Balents2010-464,Zhou2017-89,Shimizu2003-91,Pratt2011-471,Banerjee2016-15}, the fractional quantum Hall effect \cite{Stormer1999-71,de-Picciotto1997-389}, and high-temperature superconductivity \cite{Bednorz1986-64,Leggett2006-2}.

In the following, we consider low-rank PEPS (lrPEPS)
\begin{equation}\label{eq:PEPS}
	|\psi\ket = \sum_{\{\sigma_{i,j}\}}\operatorname{Contr}(A^{\sigma_{1,1}}_{1,1}\dotsb A^{\sigma_{W,L}}_{W,L})|\sigma_{1,1},\dotsc, \sigma_{W,L}\ket
\end{equation}
for a $W\times L$ square lattice with open boundary conditions. In the bulk of the system, the tensors $[A_{i,j}]^\sigma_{\mu_\text{r},\mu_\text{u},\mu_\text{l},\mu_\text{d}}$ have one physical index $\sigma$ and four virtual indices $\mu_n$, each getting contracted with a corresponding index of the tensor on a neighboring site. The tensors of the first and last rows ($j=1,L$) of the lrPEPS may have full rank. We impose a low-rank constraint \eqref{eq:lrTTNS-Ai} on all other tensors. For those in the bulk ($z=4$), it takes the form
\begin{equation}\label{eq:lrPEPS-Aij}
	A_{i,j} = \sum_{k=1}^r \tilde{a}^{k} \otimes a_\text{r}^{k} \otimes a_\text{u}^{k} \otimes a_\text{l}^{k} \otimes a_\text{d}^{k}
	\ \in\ \CC^{d\times m^{\times 4}}.
\end{equation}
\begin{figure}[t]
	\includegraphics[width=\columnwidth]{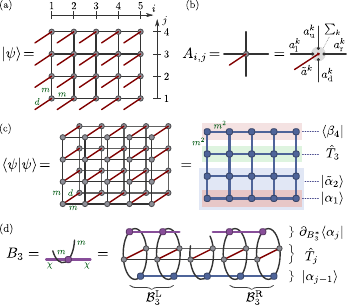}
	\caption{\textbf{Low-rank PEPS.} (a) A PEPS \eqref{eq:PEPS} with bond dimension $m$ on a $5\times 4$ square lattice. (b) lrPEPS $|\psi\ket$ are composed of rank-constrained tensors \eqref{eq:lrPEPS-Aij}. (c) Diagrammatic representations for a squared norm $\bra\psi|\psi\ket$. The rows of this tensor network can be interpreted MPOs $\hT_j$ that act on boundary MPS $|\alpha_j\ket$ and $|\beta_j\ket$.
	(d) The boundary MPS $|\alpha_j\ket$ with bond dimension $\chi$ is determined from $|\alpha_{j-1}\ket$ by an alternating least-squares minimization of the functional \eqref{eq:lrPEPS_G}. In each local optimization step, the new MPS tensor $B_i$ for column $i$ is obtained by contraction of a 1D tensor network as in the shown diagram.}
	\label{fig:PEPS_norm}
\end{figure}

The network can be optimized through imaginary time evolution or gradient-based methods. In both cases, we need to evaluate expressions similar to the squared norm, which is diagrammatically represented in Fig.~\ref{fig:PEPS_norm}c. As in the case of full-rank PEPS \cite{Verstraete2004-7,Schuch2007-98,Haferkamp2020-2}, the norm of the lrPEPS can only be evaluated approximately. To this purpose, one interprets the first and last rows of the tensor network for $\bra\psi|\psi\ket$ as boundary MPS $|\alpha_1\ket$ and $|\beta_L\ket$ with bond dimension $\chi=m^2$ and site Hilbert space dimension $m^2$. The other rows then correspond to matrix product operators (MPO) $\{\hT_j\,|\,j=2,\dots,L-1\}$ that act on the boundary MPS; see Fig.~\ref{fig:PEPS_norm}c. Generally, the cost for an exact encoding of the boundary states
\begin{equation}
	|\tilde{\alpha}_j\ket:=\hT_j\dotsb\hT_3\hT_2|\alpha_1\ket
\end{equation}
for the first $j$ rows in MPS form, increases exponentially in $j$. To avoid this, we perform an alternating least-squares optimization after every application of a row operator, to find a precise MPS approximation $|\alpha_j\ket$ of $|\tilde{\alpha}_j\ket$ with fixed bond dimension $\chi\sim m^2$. Specifically, we minimize the distance
\begin{equation}\label{eq:lrPEPS_G}
	G := \big \||\alpha_j\ket - \hT_j|\alpha_{j-1}\ket\big\|^2
\end{equation}
with respect to the MPS $|\alpha_j\ket$. As $G$ is quadratic in all MPS tensors $B_i\in\CC^{m^2\times\chi\times\chi}$ of $|\alpha_j\ket$, the optimization problem for column (vertex) $i$ reduces to the solution of a linear system of equations \cite{Schollwoeck2011-326,Jeckelmann2002-66,Verstraete2004-7}. In particular, $\partial_{\vec{B}_i^*}G=0$ is equivalent to
\begin{equation}\label{eq:lrPEPS_Bi}
	N^\eff_i \vec{B}_i = \partial_{\vec{B}_i^*}\bra\alpha_j|\hT_j|\alpha_{j-1}\ket,
\end{equation}
where we treat $B_i$ as a vector $\vec{B}_i\in\CC^{m^2\chi^2}$. Imposing orthonormality constraints \eqref{eq:ON-constr} to all $B_{i'\neq i}$ with column $i$ as the center, the effective norm matrix $N^\eff_i= \partial_{\vec{B}_i}\partial_{\vec{B}_i^*}\bra\alpha_j|\alpha_j\ket$ becomes the identity, and the left-hand side of Eq.~\eqref{eq:lrPEPS_Bi} is simply $B_i$. Figure~\ref{fig:PEPS_norm}d shows the resulting simplified version of the equation for the optimization on column $i$, where the updated tensor $B_i$ is computed by contracting the tensor network on the right-hand side.
\begin{figure}[t]
	\includegraphics[width=\columnwidth]{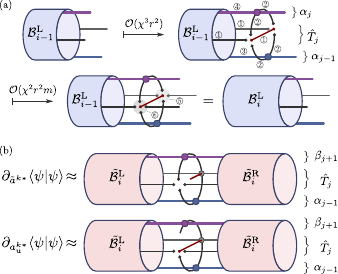}
	\caption{\textbf{Evaluation of block tensors and gradients for lrPEPS.} (a) Contraction sequence for computing block tensor $\B^\text{L}_i$ from $\B^\text{L}_{i-1}$, the column-$i$ MPS tensors of $|\alpha_{j-1}\ket$ and $|\alpha_j\ket$, and the PEPS tensor $A_{i,j}$. With the relation \eqref{eq:lrPEPSassume} between bond dimensions $m,\chi$ and the tensor rank $r$, the computation cost scales as $\O(\chi^3 r^2)\sim \O(m^6 r^2)$.
	(b) Diagrammatic representations for the derivatives \eqref{eq:lrPEPSgrad} of the (approximate) squared norm with respect to rank-one components of the PEPS tensor $A_{i,j}$ [Eq.\eqref{eq:lrPEPS-Aij}].}
	\label{fig:PEPS_contract}
\end{figure}

The low-rank property \eqref{eq:lrPEPS-Aij} of the PEPS tensors $A_{i,j}$ that comprise the row operator $\hT_j$ substantially speeds up the contraction. Let $\B^\text{L}_i$ and $\B^\text{R}_i$ denote the \emph{block tensors} defined in Fig.~\ref{fig:PEPS_norm}d, i.e., the parts of the tensor network for $\bra\alpha_j|\hT_j|\alpha_{j-1}\ket$ left of column $i$ and right of column $i$, respectively. $\B^\text{L}_i$ can be evaluated iteratively by contracting tensors from left to right. Similar to the lrTTNS algorithm, one can proceed by first contracting the rank-one components $\tilde{a}^k$ and $a_n^k$ separately and, then, performing the MPS tensor contractions and the summation over component indices [$k$ in Eq.~\eqref{eq:lrPEPS-Aij}]. An efficient contraction sequence is given in Fig.~\ref{fig:PEPS_contract}a. Assuming
\begin{equation}\label{eq:lrPEPSassume}
	d\leq m\leq r\quad\text{and}\quad \chi\sim m^2
\end{equation}
its time complexity is
\begin{equation}\label{eq:lrPEPScost}
	\O(\chi^3 r^2)\sim \O(m^6 r^2).
\end{equation}
Analogous evaluations yield $\B^\text{R}_i$ and the right-hand side of Eq.~\eqref{eq:lrPEPS_Bi}. After the update of tensor $B_i$, we move the optimization and orthonormality center $i$ to the next column. One then sweeps forth and back through $i\in\{1,\dotsc,W\}$ until the boundary MPS $|\alpha_j\ket$ has converged. The involved reorthonormalization of MPS tensors $B_i$ requires $\O(\chi^3m^2)$ operations. Progressing from row to row, we obtain all $|\alpha_j\ket$ and can similarly compute boundary MPS $|\beta_j\ket\approx \hT^\dag_j\dotsb\hT^\dag_{L-1}|\beta_L\ket$ that represent the top $L-j+1$ rows. Finally, an inner product of boundary MPS yields the lrPEPS norm
\begin{equation}
	\bra\psi|\psi\ket \approx \bra\beta_j|\alpha_{j-1}\ket\quad\forall j.
\end{equation}

A gradient-based lrPEPS groundstate optimization requires the derivatives of $\bra\psi|\psi\ket$ and $\bra\psi|\hH|\psi\ket$ with respect to the rank-one components of all tensors \eqref{eq:lrPEPS-Aij}. For tensor $A_{i,j}$, we determine the boundary MPS $|\alpha_{j-1}\ket$ and $|\beta_{j+1}\ket$ as illustrated in Fig.~\ref{fig:PEPS_norm}. The tensor networks for the gradients
\begin{subequations}\label{eq:lrPEPSgrad}
\begin{align}
	\partial_{\tilde{a}^{k*}}\bra\psi|\psi\ket &\approx \partial_{\tilde{a}^{k*}}\bra\beta_{j+1}|\hT_j|\alpha_{j-1}\ket,\\
	\partial_{a^{k*}_n}\bra\psi|\psi\ket &\approx \partial_{a^{k*}_n}\bra\beta_{j+1}|\hT_j|\alpha_{j-1}\ket
\end{align}
\end{subequations}
of the norm with respect to the rank-one components $\tilde{a}^k\in\CC^d$ and $a_\text{r}^{k},a_\text{u}^{k},a_\text{l}^{k},a_\text{d}^{k}\in\CC^m$ of $A_{i,j}$ 
are shown in Fig.~\ref{fig:PEPS_contract}b. The corresponding block tensors $\tilde{\B}^\text{L}_i$ and $\tilde{\B}^\text{R}_i$ are defined as in Fig.~\ref{fig:PEPS_norm}d, with MPS $\bra\alpha_j|$ replaced by $\bra\beta_{j+1}|$.

The time complexity for one optimization step in this lrPEPS algorithm scales as $\O(\chi^3 r^2)\sim \O(m^6r^2)$. Assuming that we can choose $r \sim m$ as in the lrTTNS simulations, the $\O(m^8)$ cost is substantially below the $\O(m^{10})$ cost for full-rank PEPS \cite{Murg2007-75}.

\section{Discussion}
\emph{Generalization to fermions and iPEPS.} --
A big strength of TNS techniques is that they are also applicable for frustrated quantum magnets and fermionic systems \cite{Barthel2009-80,Corboz2009-80,Pineda2009_05,Kraus2009_04,Corboz2009_04}, where quantum Monte Carlo is hampered by the negative-sign problem \cite{Loh1990-41,Troyer2005}. 
Following the formulation in Ref.~\cite{Barthel2009-80}, it is straightforward to see that the low-rank TNS approach can be adapted to fermionic systems. In particular, when choosing the rank-one components of the rank-constrained tensors [Eqs.~\eqref{eq:CPD}, \eqref{eq:lrTTNS-Ai}, and \eqref{eq:lrPEPS-Aij}] as elements of a reduced fermionic Fock space of dimension $m$ or $d$, respectively, the low-rank TNS $|\psi\ket$ is automatically anti-symmetric under particle permutations. Imposing that the rank-one components have either even or odd particle number parity, the discussed tensor contractions remain basically the same, involving only some additional sign factors \cite{Barthel2009-80}. Note also, that the approach for lrPEPS, described in Sec.~\ref{sec:lrPEPS} can easily be adapted to infinite PEPS (iPEPS), where one simulates directly in the thermodynamic limit \cite{Jordan2008-101,Orus2009_05,Corboz2016-94}.

\emph{TNS with tensors in Tucker format.} --
The Tucker decomposition \cite{Tucker1966-31,Hitchcock1927-6} is a generalization of the canonical polyadic decomposition \eqref{eq:CPD}, where an order-$z$ tensor $A_i$ is composed of a small order-$z$ core tensor and $z$ matrices (instead of the $z$ rank-one components). The canonical polyadic decomposition, corresponds to diagonal core tensor. Using a Tucker format for the TNS tensors, one can seamlessly tune from the low-rank TNS discussed in this work to full-rank TNS, increasing both the expressiveness of the TNS and the computation costs. It remains to be tested whether this might result in even more efficient simulations.

\emph{Conclusion.} -- 
The proposed low-rank TNS are a promising approach for more efficient simulations of strongly-correlated quantum many-body systems and machine learning on large data sets. We discussed in some detail the cases of TTNS and PEPS, where the rank constraints reduce the computation costs to $\O(z(m^2r+r^2m))$ and $\O(m^6r^2)$, respectively. This is a considerable improvement over the traditional counterparts, $\O(m^{z+1})$ or $\O(m^{2z-1})$ for TTNS and $\O(m^{10})$ for 2D PEPS. In lrTTNS simulations for the quantum XXZ model, we found that setting the rank $r\sim m$ already gives high-accuracy ground state energies, and lrTTNS obtained more precise ground states than standard TTNS with considerably fewer parameters.

\begin{acknowledgments}
We gratefully acknowledge discussions with participants of the IPAM program \emph{``Tensor methods and emerging applications to the physical and data sciences''} (2021) and
support through U.S.\ Department of Energy grant DE-SC0019449.
\end{acknowledgments}

\end{document}